# StorySpace: Technology Supporting Reflection, Expression and Discourse in Classroom Narrative

Benjamin Watson[1], Janet Kim[2], Tim Mceneany[2], Claudia Hindo[3], Tom Moher[2], Louis Gomez[3] & Stephen Fransen[4]

[1]Dept. Computer Science
Northwestern University

[2]Dept. Computer Science
Univ. Illinois Chicago

[3]Program of Learning Sciences
Northwestern University

[4]Clemente High School
Chicago, IL

*Classroom Narrative*

As former and perhaps still current students, we can all remember essays we've written, dioramas we've built and posters we've pasted together. All these are forms of classroom narrative, a basic and important activity used by instructors to reinforce learning. The very fact that we can still remember these narratives decades later shows how successful this activity can be.

The power of classroom narrative lies in the wide range of abilities it engages. Most importantly, narrative engages the student's reflective and interpretive faculties. In producing a narrative artifact, students must internalize the classroom topic and most often also adopt a certain point of view about it. Narratives can also engage a wide range of expressive abilities. For example, writing an essay requires composition and language skills. Building a diorama is a much more tangible experience, in which students must select a key scene or concept from the classroom topic, and then physically illustrate it using inexpensive building blocks or materials. Composing a poster taps into graphic and editorial skills, with students searching for good illustrations of key concepts in other media (e.g. National Geographic), and then arranging them into an informative and illustrative whole. Finally, narratives sometimes exercise student ability in presentation and discourse. When narratives are group projects or are presented before the class or at school fairs, students must collaborate with fellow group members to select and shape narrative content, then introduce and explain the result to fellow students, parents and the wider community.

*Designing Technology for Classroom Narrative*

The StorySpace project studies the role new interface technologies [1][2] might play in high school education. Unfortunately, technology often seizes center stage in the classroom, becoming itself the topic of instruction. We argue that learning in general and learning about technology in particular will be most successful when technology is used in the service of learning, rather than learning in the service of technology. Certainly outside of the classroom, technology is rarely an end in itself – application gives technology its power. The classroom should be no different.

With this approach in mind, StorySpace is specifically designed to support and enhance classroom narrative, already a well established classroom activity. StorySpace strives to achieve this through adherence to the following design goals:

> *Trigger student reflection and interpretation.* The narrative medium created by StorySpace should be capable of representing the topic of classroom discussion and learning in all its complexity. In building their representation, the students will then be confronted with that same complexity. The medium should also itself be exciting and compelling, making classroom narrative interesting and fun.

> *Accommodate individual student expression.* Each student's perspective on the classroom topic should be easily captured and represented in the narrative artifact they produce. Students have a wide range of personalities and come from various ethnic, cultural and educational backgrounds. To reach all of them, the StorySpace medium should therefore be more than textual and include easily learned physical, visual and auditory elements. Students should initially be convinced that they can use the medium, and fascinated by the opportunity to improve the artifact they have already built.

> *Encourage student discourse.* Many classroom narratives take the form of largely one-way, one-to-one communications between students and their teacher. Many researchers believe that learning increases when narratives instead become a true discourse, with students engaging in a two-way conversation with not only their teacher, but also their peers and community. StorySpace should allow and encourage the two-way, many-to-many forms of communication that are the hallmark of rich classroom discourse and discussion.

By adhering to these goals, StorySpace will be not only an effective tool for classroom narrative, but also an effective tool for learning about technology. The best artists are not only very knowledgeable about their subject, but inevitably also experts in their medium. Similarly, students building a compelling narrative in StorySpace will become very familiar with its technology. This will be particularly true if they find the StorySpace medium rich and exciting, if they see clear opportunities to express themselves, and if they are addressing their peers and community in addition to their teacher.

*The StorySpace Design*

StorySpace enables the construction and display of a new form of narrative that is simultaneously digital and physical, a form that is probably most similar to the traditional paper poster. Narratives are represented visually on a two-dimensional tabletop board, and physically with three-dimensional tokens. Audience members can not only experience the narrative passively, but can also interact with the narrative by manipulating the tokens, changing the visual representation. With many tokens around the table providing multiple points of access for the audience, collaborative group interaction is easily supported. Audience members can also alter the narrative itself, leaving annotations and "graffiti" behind them.

StorySpace supports student design of the narrative itself as a special form of narrative alteration. Students collect and build media using the web and a variety of portable devices including PDAs, cell phones, digital recorders, and digital cameras. These media (including imagery, sound, movie clips) are loaded onto a blank board through an attached PC. Once on the board, media may be manipulated using predefined tokens for moving, removing, resizing, zooming, playing and stopping media. An additional token provides an undo operation. These manipulations are made by placing the appropriate token on a piece of media, and if appropriate moving that token (e.g. to indicate the new location of the piece). Manipulations may also be recorded as macros, enabling higher level interaction semantics such as playback of a movie clip that first expands to fill the entire board.

Figure 1: A close up of the StorySpace board in its design mode.

We constructed StorySpace's board from one or more interactive chessboard substrates built by DGT. These inexpensive substrates are designed to attach to a PC and track the location of the pieces on a chessboard. Pieces are identified with radio frequency (RF) tags. We cover the substrates with a white laminated surface, reflective enough for projection and durable enough for interaction. A lighting stand and attached custom cantilever mount holds a digital projector over the surface. We embed the tags into the StorySpace tokens. The design tokens themselves are custom fabricated using a rapid prototyping three-dimensional printer, with shapes suggesting their function and receptacles for the RF tags (students can also design custom tokens specialized to each narrative). A host PC accepts input from the substrate and sends board output to the projector.

Figure 2: The StorySpace table, projector and stand.

StorySpace was inspired by research in both tangible and ubiquitous interfaces. Tangible user interfaces [1] embed the human-computer interface in physical objects, exploiting the natural spatial abilities of their users and opening not only visual but physical paths of communication. The StorySpace board has much in common with these interfaces. Ubiquitous [2] user interfaces stress the embedding of computational devices into our everyday lives, specializing those devices to make their use minimally disruptive, and often miniaturizing them to make them accessible at all times. StorySpace's media collection devices exploit this technology to take learning out of the classroom and into students' communities and personal lives.

This design adheres quite well to our goals for a successful tool for classroom narrative. Student reflection and interpretation is triggered with the use of digital imagery, video and audio as source material. All of these are rich, dynamic media that today's students know and love. StorySpace allows students to sample and combine these media in many modalities, like much of the art in this postmodern age. Individual student expression is accommodated through the use of an easily manipulated tangible interface, lowering barriers to use by those with different cultural and educational backgrounds and capturing the expressive strengths of visually and tactilely oriented students. Student dialogue and discourse are encouraged with collaborative tabletop interface that affords discussion in design and display, ubiquitous components that take media collection into students social and community lives, and a nonlinear, self-referential medium that accommodates many perspectives.

*StorySpace in the Classroom*

Of course, the real test of any design is its use and evaluation in practice. We are currently integrating StorySpace into the curriculum of an English literature class at Clemente High School in Chicago.

In a previous class, students used Apple's iMovie to construct narratives about Edwards' "Sinners in the Eyes of an Angry God". The results were quite compelling. Students not only learned about technology of digital video and search (for imagery), but also demonstrated a deep metaphorical understanding of the text. Student effort and standards of quality were raised by the highly visible display of their work to peers.

A pilot group of these same students is now using StorySpace to build in-class narratives about Shakespeare's MacBeth. Our focus at this early stage is on narrative design, as opposed to ubiquitous media collection or display to an audience. The students have already received instruction in the use of StorySpace, and are now collecting media in the classroom. The students are clearly excited and intrigued by the tool. Experience to date has already resulted in some design changes, including a simplified set of design manipulations, streamlined procedures for loading media onto the board and the use of multiple projectors supporting wider classroom discussion.

Figure 3: Students learn how to use StorySpace.

After they have built their narrative, our pilot group of students will answer questions requesting suggestions they might have for improving the technology, the strengths of StorySpace as compared to text and other digital media, and probing their engagement in a discourse with their peers about both the narrative content and technology. The students will then present their narrative to the rest of the class, giving them their perspective on MacBeth and educating them in the use of StorySpace. At this point, the entire class will be split into groups to build a number of narratives examining the characters, themes, and ancient as well as modern interpretations in detail.

*The Future of StorySpace*

Beyond these immediate goals, we plan to expand the use of StorySpace at Clemente High School to study its effectiveness in display to an audience of peers. We also hope to create new connections between the school and its community by using StorySpace to display classroom narratives in neighborhood centers, and by sending students equipped with ubiquitous media collectors into the streets and blocks around their homes.

The StorySpace board is unique and currently still fragile resource, and loading media onto it can be tedious. In the long term we will have to improve the board and accommodate its shortcomings in pedagogy. We also anticipate improving the utility of StorySpace by using it in other classes centered around the physical and social sciences, for which we will have to develop matching pedagogies. Ultimately, we hope StorySpace can eliminate some of the digital divide by giving students of diverse backgrounds a new perspective and comfortable mastery of technology.

*Acknowledgements*

This work is supported by NSF grant IIS-0112937. Our thanks to Brian Dennis for many valuable discussions, and especially to the administration and staff at Clemente High School for their enthusiasm and flexibility in accommodating this work.

Figure 1:

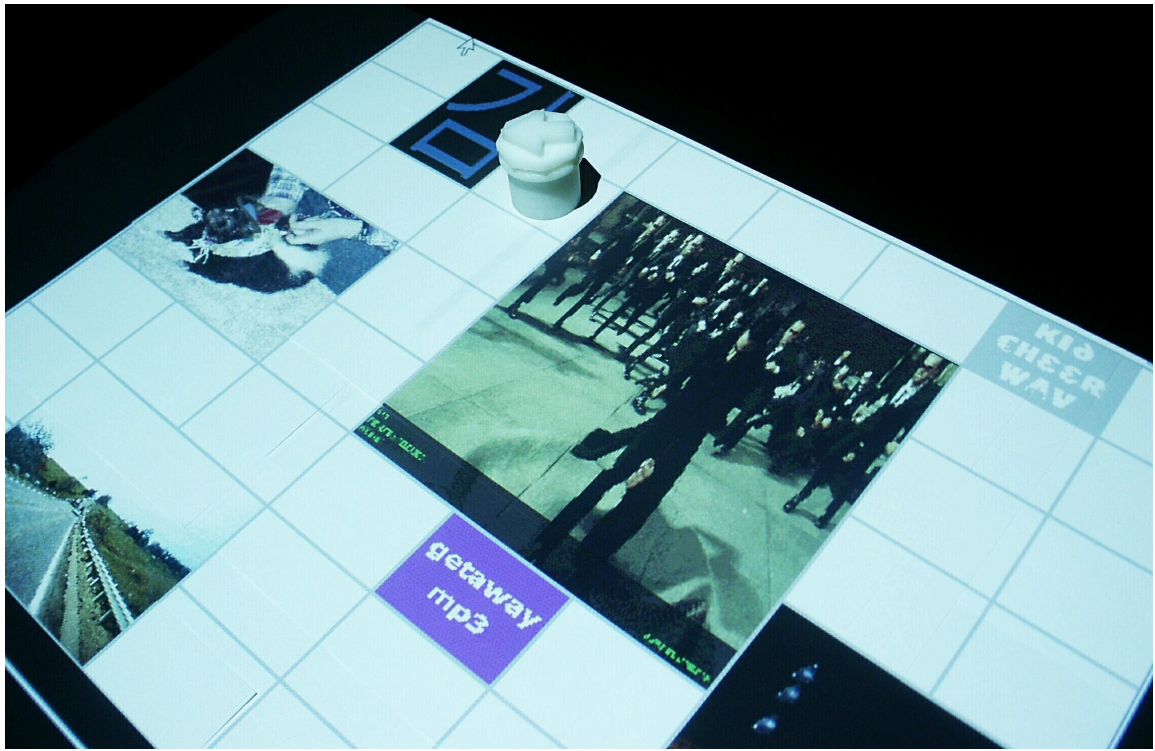

Figure 2:

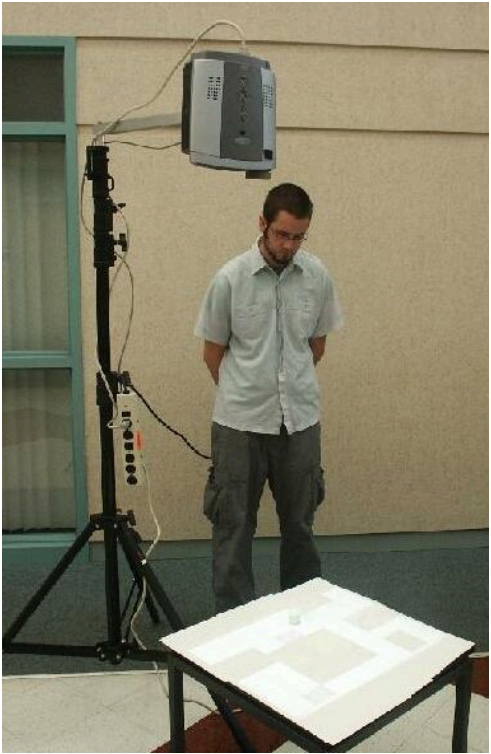

Figure 3:

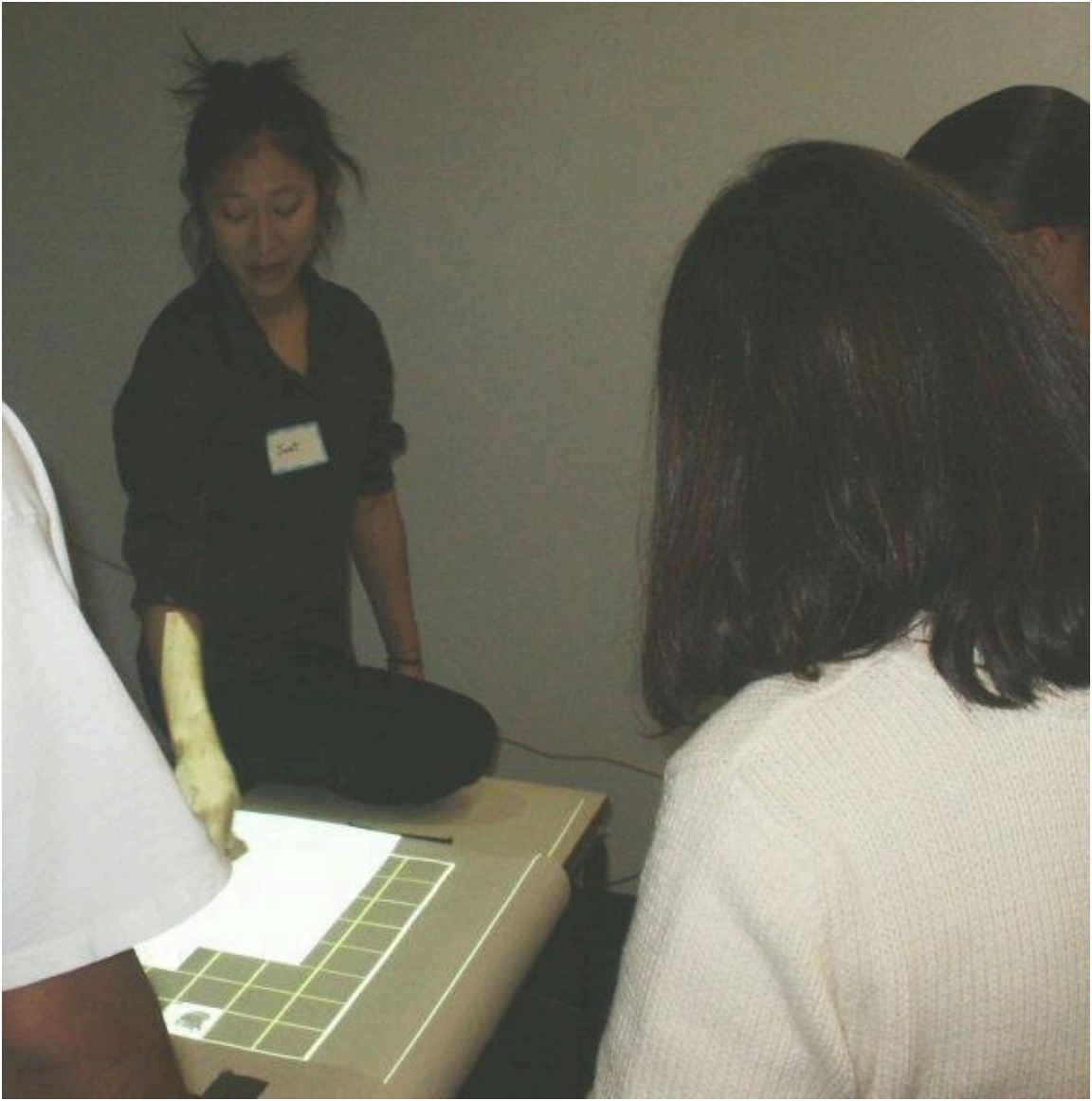